\documentclass[prl,twocolumn,showpacs,floatfix,preprintnumbers,amsmath,amssymb,superscriptaddress]{revtex4}

\usepackage{graphicx}
\usepackage{dcolumn}
\usepackage{bm}
\usepackage{color}
\usepackage{color}
\usepackage[latin1]{inputenc}
\usepackage[urlcolor=blue]{hyperref}
\hypersetup{backref, colorlinks=true, linkcolor=blue, citecolor=blue}

\begin{document}

\title{{Exotic Magnetic and Electronic Properties of Layered CrI$_{3}$ Single Crystals Under High Pressure }}

\author{Anirudha Ghosh}
\affiliation{Uppsala University, Department of Physics and Astronomy, Box 516, SE-751 20 Uppsala, Sweden}

\author{D. Singh}
\affiliation{Uppsala University, Department of Physics and Astronomy, Box 516, SE-751 20 Uppsala, Sweden}

\author{Qingge Mu}
\affiliation{Max Planck Institute for Chemical Physics of Solids, D-01187 Dresden, Germany}

\author{Y. Kvashnin}
\affiliation{Uppsala University, Department of Physics and Astronomy, Box 516, SE-751 20 Uppsala, Sweden}

\author{G. Haider}
\affiliation{J. Heyrovsky Institute of Physical Chemistry of the Czech Academy of Sciences, Dolej$\check{s}$kova 2155, 182 23 Prague, Czech Republic}

\author{M. Jonak}
\affiliation{Kirchhoff Institute of Physics, Heidelberg University, Heidelberg, Germany}

\author{D. Chareev}
\affiliation{National University of Science and Technology "MISiS", Moscow, 119049, Russia}

\author{T. Aramaki}
\affiliation{Graduate School of Engineering, Kyushu Institute of Technology, Fukuoka 804-8550, Japan}

\author{S.A. Medvedev}
\affiliation{Max Planck Institute for Chemical Physics of Solids, D-01187 Dresden, Germany}

\author{R. Klingeler}
\affiliation{Kirchhoff Institute of Physics, Heidelberg University, Heidelberg, Germany}

\author{M. Mito}
\affiliation{Graduate School of Engineering, Kyushu Institute of Technology, Fukuoka 804-8550, Japan}

\author{E. H. Abdul-Hafidh}
\affiliation{Physics Department, Faculty of Science at Taibah University-Yanbu, King Khalid Rd. Al Amoedi, 46423, Yanbu El-Bahr, 51000, Saudi Arabia}

\author{J. Vejpravova}
\affiliation{Department of Condensed Matter Physics, Faculty of Mathematics and Physics, Charles University, Ke Karlovu 5, 121 16 Prague 2, Czech Republic}

\author{M. Kalb$\grave{a}$$\check{c}$}
\affiliation{J. Heyrovsky Institute of Physical Chemistry of the Czech Academy of Sciences, Dolej$\check{s}$kova 2155, 182 23 Prague, Czech Republic}

\author{R. Ahuja}
\affiliation{Uppsala University, Department of Physics and Astronomy, Box 516, SE-751 20 Uppsala, Sweden}

\author{Olle Eriksson}
\affiliation{Uppsala University, Department of Physics and Astronomy, Box 516, SE-751 20 Uppsala, Sweden}
\affiliation{School of Science and Technology, Örebro University, SE-701 82 Örebro, Sweden}

\author{Mahmoud Abdel-Hafiez}
\email{mahmoud.hafiez@physics.uu.se}
\affiliation{Uppsala University, Department of Physics and Astronomy, Box 516, SE-751 20 Uppsala, Sweden}

\date{\today}

\begin{abstract}
Through advanced experimental techniques on CrI$_{3}$ single crystals, we derive a previously not discussed pressure-temperature phase diagram. We find that $T_{c}$ increases to $\sim$ 66\,K with pressure up to $\sim$ 3\,GPa followed by a decrease to $\sim$ 10\,K at 21.2\,GPa. The experimental results are reproduced by theoretical calculations based on density functional theory where electron-electron interactions are treated by a static on-site Hubbard U on Cr 3$d$ orbitals. The origin of the pressure induced reduction of the ordering temperature is associated with a decrease of the calculated bond angle, from 95$^{\circ}$ at ambient pressure to $\sim$ 85$^{\circ}$ at 25\,GPa. Above 22\,GPa, the magnetically ordered state is essentially quenched, possibly driving the system to a Kitaev spin-liquid state at low temperature, thereby opening up the possibility of further exploration of long-range quantum entanglement between spins. The pressure-induced semiconductor-to-metal phase transition was revealed by high-pressure resistivity that is accompanied by a transition from a robust ferromagnetic state to gradually more dominating anti-ferromagnetic interactions and was consistent with theoretical modeling.

\end{abstract}


\maketitle

Two-dimensional (2D) van der Waals (vdW) materials offer a plethora of functional properties that are not only of fundamental interest but are essential for the development of new technological applications\cite{1,2,3}. Layered chromium trihalides emerged as potential 2D vdW materials with unique layer-dependent magnetic properties\cite{4,5}. Understanding the long-range magnetic order in these 2D materials is an intriguing subject of widespread research. According to the Mermin-Wagner theorem\cite{6}, it is strongly suppressed in a 2D isotropic Heisenberg system due to spin fluctuations at any finite temperature. Here, the magnetocrystalline anisotropy (MCA) comes to the rescue and stabilizes the long-range magnetic order in CrI$_{3}$\cite{7,8,9,10,11}. Several experiments and theories have been put forward, reflecting the role of I-5$p$ state spin-orbit coupling (SOC) strength in the origin of the MCA through the Cr 3$d$-I 5$p$-Cr 3$d$ superexchange interaction. This interaction is mediated by I-ions octahedrally coordinating with Cr ions with an in-plane Cr-I-Cr bond angle of 95°. According to the Goodenough-Kanamori-Anderson (GKA) rule, the magnetic interaction is primarily ferromagnetic (FM) when the metal-ligand-metal bond angle is 90$^{\circ}$\cite{12}. Therefore, the Cr-I-Cr bond angle plays a significant role in understanding the long-range FM superexchange interaction\cite{13}, as well as the MCA. Further modification in the bond angle should have a substantial impact on the electronic, magnetic, and transport properties of CrI$_{3}$. Theoretical calculations of a favorable magnetic ground state in monolayer CrI$_{3}$ are not too conclusive, as they suggest both FM\cite{14} and anti-ferromagnetic (AFM)\cite{15} ground states under lattice compression. However, a more recent calculation suggests that depending on the anisotropy of the in-plane lattice strains, both these ground states can co-exist with a possibility of a complete quenching of the magnetic order at a critical isotropic lattice compression\cite{16}. Experimentally, an increase in the FM $T_{c}$ with pressure was attributed to a possible decrease in the Cr-I-Cr bond angle towards 90$^{\circ}$\cite{17}. However, with a maximum pressure of only 1\,GPa, the variation of $T_{c}$ in a broader range of pressures eluded experimental detection. This prohibits an experimental analysis of the possible decrease in the Cr-I-Cr bond angle below 90$^{\circ}$, and its influence on the strength of the Cr-Cr superexchange interaction and $T_{c}$.

\begin{figure*}[!t]
\includegraphics[width=1.8\columnwidth]{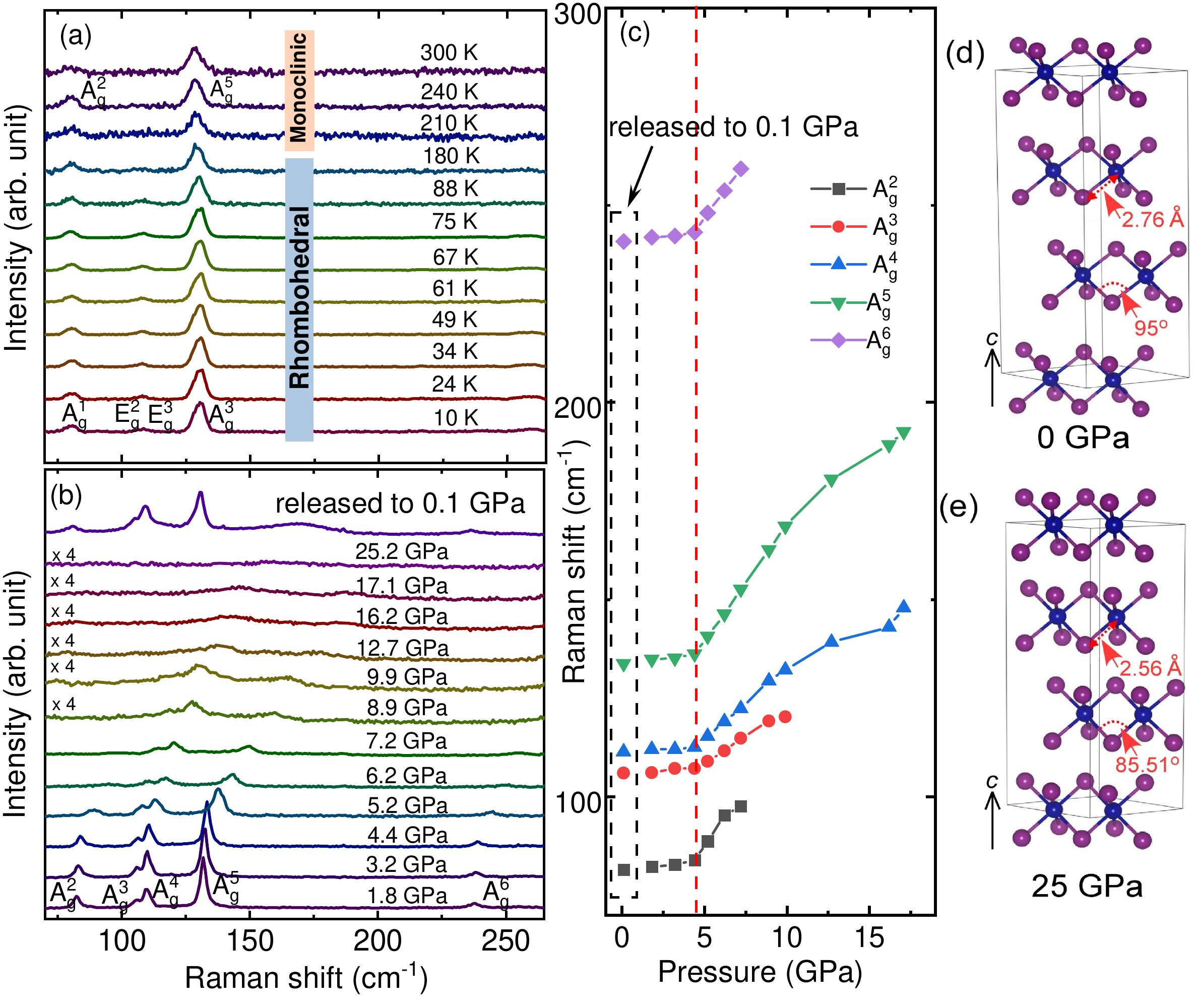}
\caption{(a) Temperature-dependent Raman spectra measured with circularly polarized light in parallel polarization configuration (see SI for detailed analysis) (b) Pressure-dependent Raman spectra at 300\,K measured at parallel polarization configuration. (c) Raman shift of all the phonon modes with pressure obtained from the spectra presented in panel (b). Red dashed line shows the pressure above which blue shift and smearing of peaks occur. Calculated structure (d) at 0\,GPa and (e) at 25\,GPa.}
\label{fig:calc}
\end{figure*}

The experiments reported here go up to $\sim$ 40\,GPa, and offer completely new possibilities to tune the $T_{c}$, phonon dispersions, electronic, magnetic, and magneto-transport properties of CrI$_{3}$ single crystals. Overall, understanding the correlation between the bond-angle and various magnetic and electronic properties is questionable and elusive. Here, we provide a comprehensive answer to this question for CrI$_{3}$ through combined complementary experimental and theoretical studies\cite{18}. We observe, a hitherto unexplored, pressure-induced semiconductor-to-metal (SM) phase transition and possibly a Kitaev spin-liquid (KSL) phases in CrI$_{3}$ single crystal samples.

\begin{figure}[tbp]
\includegraphics[width=20pc,clip]{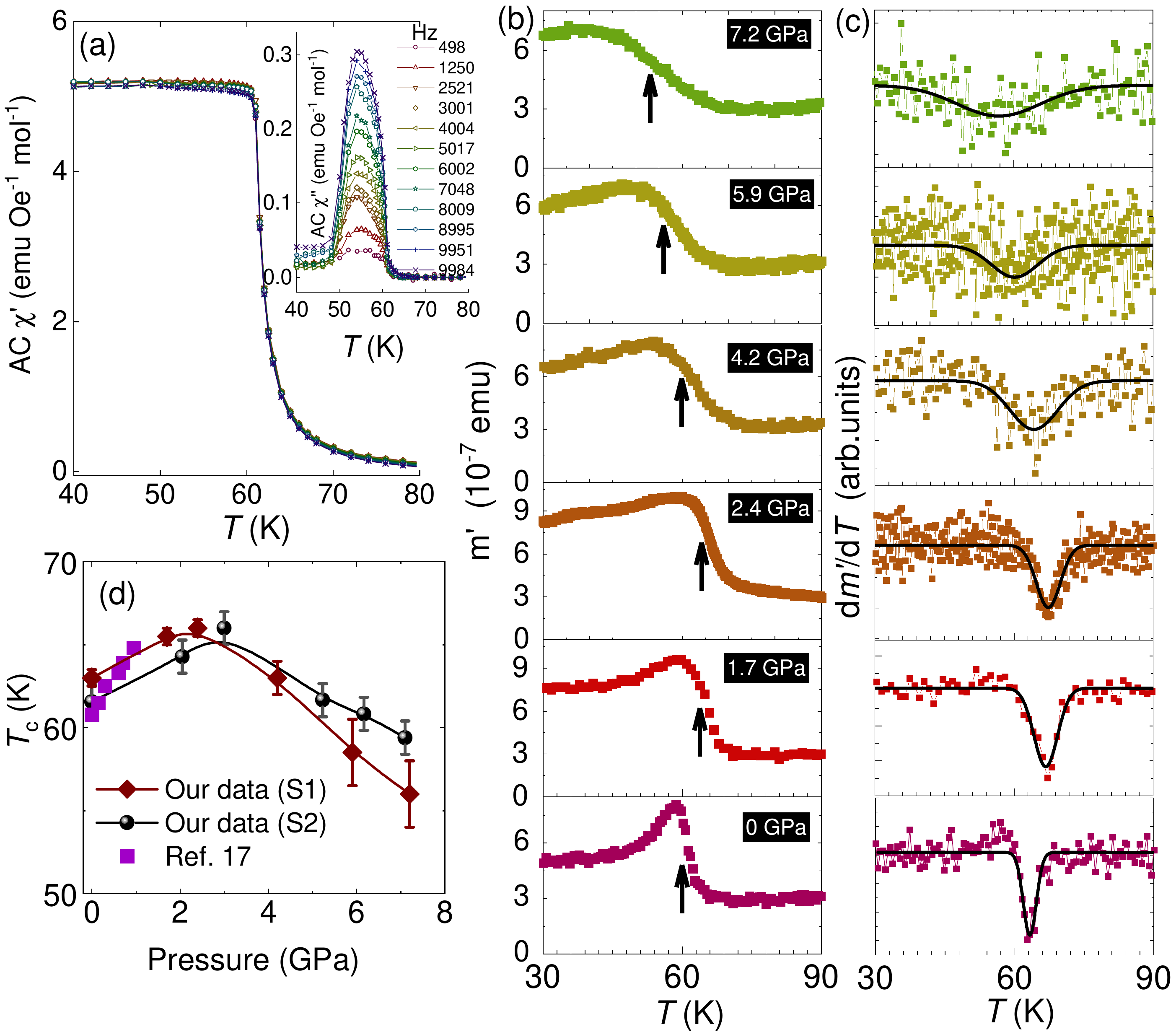}
\caption{(a) Real and imaginary (inset) part of ac susceptibility as a function of frequency of the oscillating field, $H_{AC}$ = 2\,Oe. (b) Pressure dependent real part of the ac susceptibility plots at a magnetic field of $H_{AC}$ = 3.86\,Oe oscillating at a frequency of $f$ = 10\,Hz of crystal (S2). (c) $dm^{\prime}/dT$ plot of the data shown in (b), stacked over one another for clarity. Solid black lines represent Gaussian fit to find the minimum which represent $T_{c}$ and in agreement with the arrows in (b) and plotted in (d). The error bars represent the computational error in the Gaussian fit. Data points from\cite{17} are shown for comparison.}
\end{figure}


Temperature-dependent Raman spectra at ambient pressure shows a FM to paramagnetic ($T_{c}$) and rhombohedral ($R\bar{3}$) to monoclinic ($C2/m$) structural ($T_{s}$) transitions at $\sim$ 60\,K and $\sim$ 210\,K\cite{19}, respectively (see Fig.\,1(a) and for detailed analysis of Raman experiment see SI\cite{18}). The obtained ambient pressure Raman spectra are in excellent agreement with theoretically obtained spectra of perfect crystals, suggesting a high crystalline purity of the synthesized material\cite{20,21,22}. Pressure-dependent Raman spectra reported here, see Fig.\,1(b), presents only small variation in the optical phonon frequencies up to 4.4\,GPa. While above 4.4\,GPa in Fig.\,1(c), a sizable blue shift with a significant decrease in intensities of all the optical phonon frequencies is observed. Additionally, above this pressure, the distinctive phonon spectral features begin to smear out into broad features, and the $A_{g}^{2}$ and $A_{g}^{6}$ phonon modes are gradually suppressed and disappear above 7.2\,GPa. Above 17.1\,GPa, the rest of all the phonon modes are suppressed, which is indicative of a pressure induced deformation / distortion of the lattice. The phonon modes, re-appear when the pressure is released to near-ambient conditions. However, a broad feature near 175\,cm$^{-1}$ appears when pressure is released to 0.1\,GPa, which is reminiscent of a high-pressure phase and suggest that the structural distortions in the pressure cycle are not perfectly reversible\cite{18}. The broadening and suppression of the vibration modes, might be indicative of modulation in the Cr-I-Cr bond angle as well as an indication towards metallization. Our density functional theory (DFT) calculations reported here do however show a distinct decrease in the bond angle and Cr-I bond length with pressure, see Figs.\,1(d) and (e), without any indication of structural transition (see Fig.\,S6)\cite{18}.

The optimized lattice parameters at zero pressure are calculated from DFT to be $a$=$b$=6.886\,${\AA}$ and $c$ = 19.820\,${\AA}$, which are consistent with the previous results\cite{19}. Since the FM superexchange interaction strength is associated with the Cr-I-Cr bond angle, a modification in the same is expected to influence the magnetic properties. To analyze this further, we have carried out magnetization experiments at various external pressures. At ambient conditions, Fig.\,2(a) shows no frequency-dependent shift in the $T_{c}$ for both $\chi^{\prime}$ and $\chi^{\prime\prime}$, signifying a non-spin-glass behavior\cite{23}, (see also Fig.\,S7). Figures\,2(b) and (c) summarize the variation of the real part of ac magnetization with pressure up to 7.2\,GPa and its first derivative, respectively. $T_{c}$ of two samples is obtained from the peaks of the Gaussian fit of the $dm^{\prime}/dT$ plots and are plotted in Fig.\,2(d). An increase in broadening of the Gaussian shaped peaks in Fig.\,2(c) with pressure illustrates that the phase transition is much less distinct with atomic moments that disorder over a much wider temperature range. The overall variation in $T_{c}$ can be attributed to the change in the Cr-I-Cr superexchange interaction owing to a variation in the Cr-I-Cr bond angle with pressure as shown in Figs.\,1(d)-(e). In agreement with the observed variation of $T_{c}$, our theoretical studies of the crystal structure imply a decrease of the bond angle towards 90$^{\circ}$ for pressures smaller than 4\,GPa. The calculated bond angle is about 90$^{\circ}$ at 4\,GPa, close to the pressure where $T_{c}$ is maximum, which is in agreement with the Goodenough-Kanamori rule\cite{12}. From the general line-shape of $m^{\prime}$ data in Fig.\,2(b)\cite{24}, it is evident that a magnetically ordered state is the ground state up to the maximum measured pressure of 7.2\,GPa. This is partially consistent with earlier theoretical calculations where a FM ground state was predicted even at a high compressive lattice strain\cite{14}.

To understand the magneto-transport properties and the fate of magnetic ordering at high pressures beyond the numerous studies in magnetoresistance (MR)\cite{25,26}, we have carried out MR measurements at different high pressures. Figure 3(a-c) illustrates the field-dependent MR at various temperatures and 21.2, 24.0, and 37.8\,GPa, respectively. A negative MR can be explained as a consequence of suppressed spin-spin scattering in a FM ordered state\cite{27}. Since the current is applied along the sample plane, the incoming electrons' spin will always experience a lattice with parallel spin configuration (due to intra-layer ferromagnetism) below $T_{c}$. MR can increase (positive MR) if the in-plane spins in the lattice are disordered (i.e. not primarily FM). At 21.2\,GPa, the negative MR fairly saturates at high magnetic fields and temperature below 20\,K, with the minimum at 10\,K, indicating that the onset of the FM ordered state should lie near 10\,K. At 24\,GPa in Fig.\,3(b), no saturation in the negative MR is observed even up to the lowest measured temperature. However, an initial negative downturn of low-temperature curves suggests that there could be a few FM-ordered spins. Interestingly at 37.8\,GPa in Fig.\,3(c), we do not observe any negative MR, therefore indicating a substantial quenching of FM ordering down to the minimum measured temperature, 2\,K. A positive, non-saturating MR, even at 9\,T magnetic field, stems from the enhanced spin-spin scattering possibly due to the disordered spins. Since the FM $T_{c}$ also decreases with increase in pressure above 3\,GPa, the disordered spin state is not limited to the surface, rather it's a bulk property. Our present results show the emergence of such a magnetically quenched state in the region above 22\,GPa. This pressure is equivalent to an isotropic compressive lattice strain of 14$\%$  as compared to that of 5$\%$ in monolayer CrI$_{3}$\cite{16}. Such a magnetically quenched state in an in-plane honeycomb lattice, like CrI$_{3}$, is intriguing in the context of the topological KSL phase\cite{28,29,30}. In Fig.\,3(d), our data shows a change in the slope of the resistivity curves above 21.2\,GPa, which resembles a transition to a metallic state above 21.2\,GPa (with the hole as the majority charge carrier shown in Fig.\,S8)\cite{18}), in agreement with the pressure-induced broadening of Raman peaks in Fig.\,1(b) that we interpret are due to metallization. These measurements reveal that the quenching of magnetic ordering concurrently evolves in the metallic state.

\begin{figure}[tbp]
\includegraphics[width=20pc,clip]{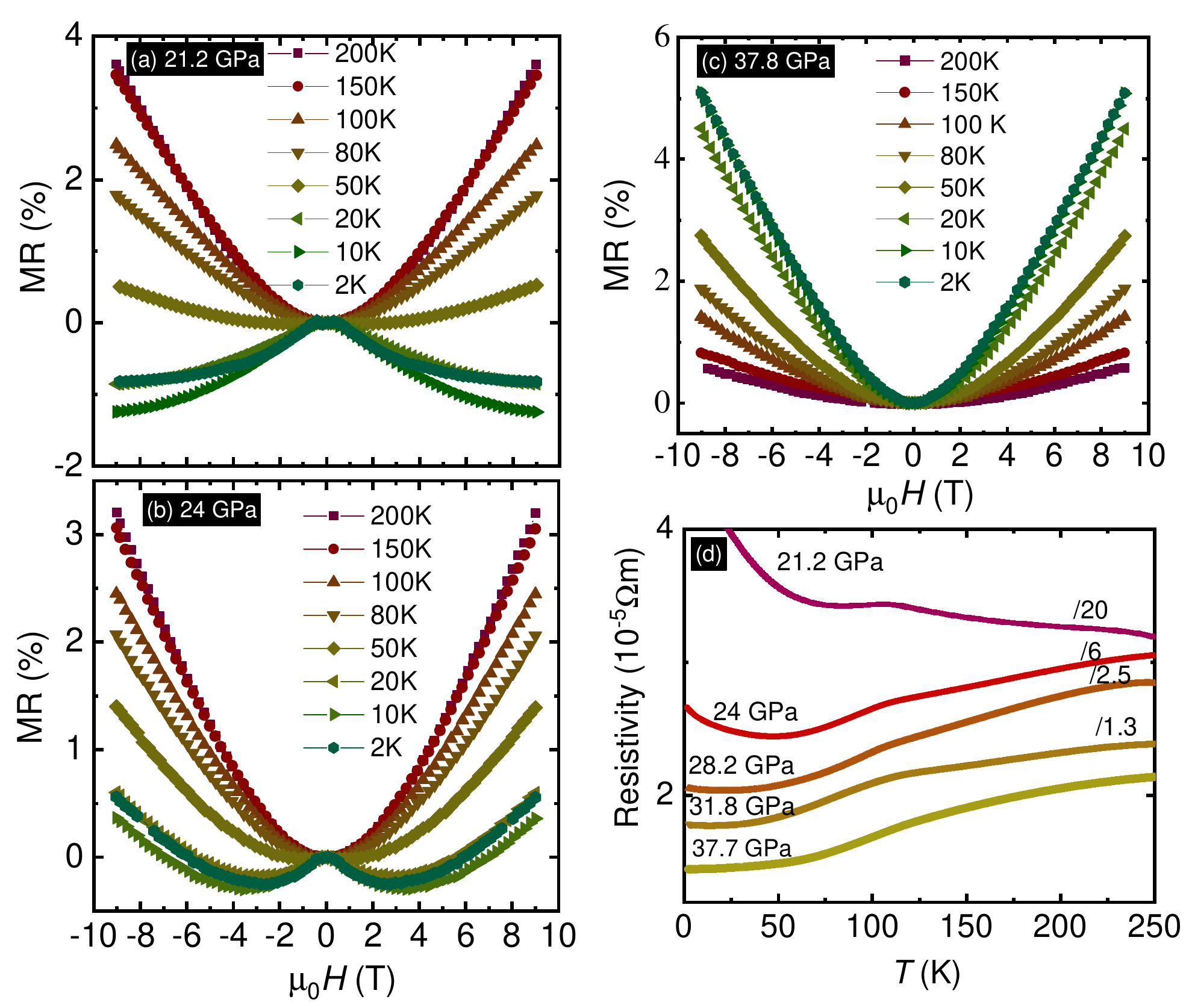}
\caption{Field-dependent MR under 21.2\,GPa (a), 24\,GPa (b), and 37.8\,GPa (c). (d) Temperature-dependent resistivity at high pressure, illustrating a semiconductor to metal transition above 21.2\,GPa.}
\end{figure}

\begin{figure}[tbp]
\includegraphics[width=20pc,clip]{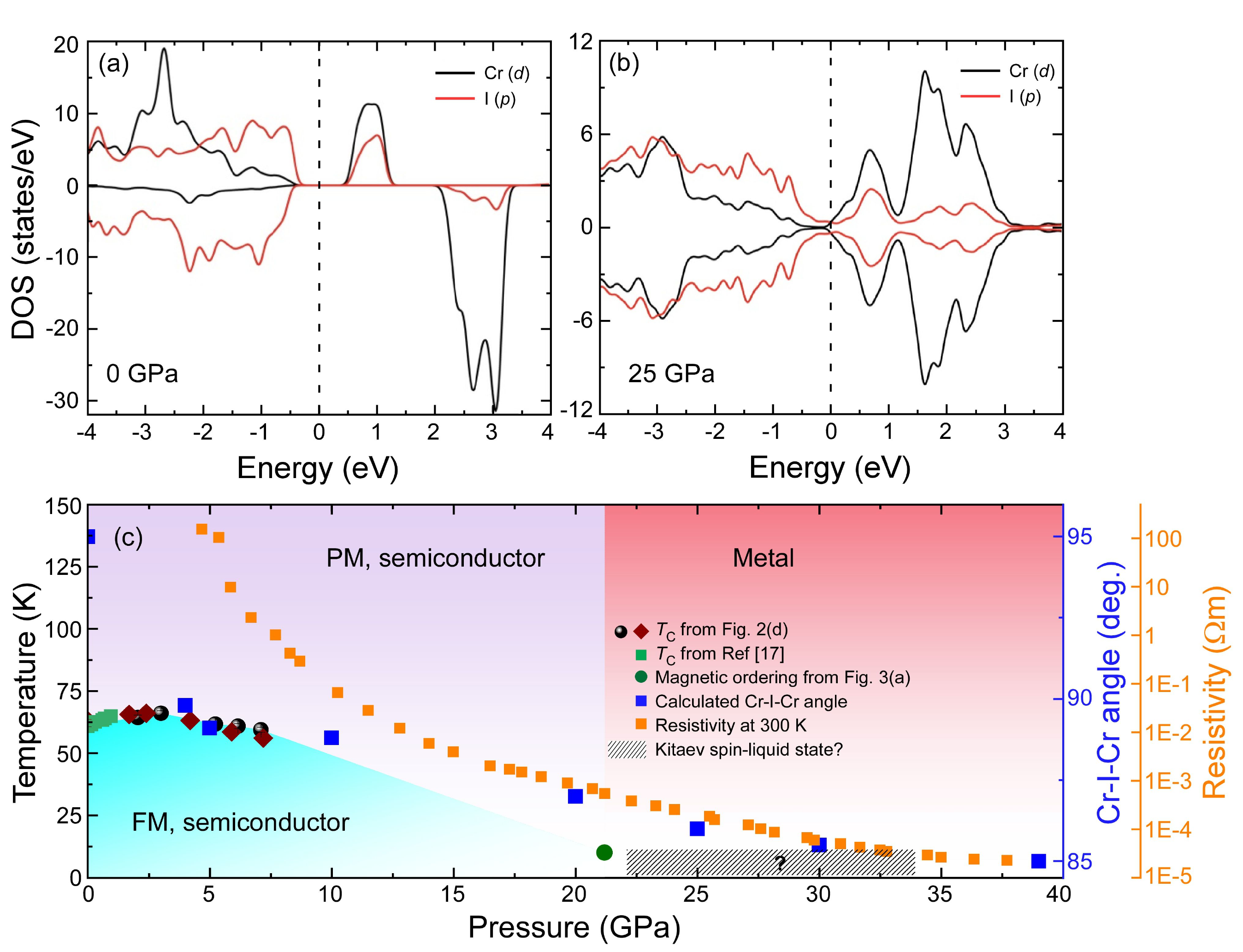}
\caption{Calculated density of states (DOS) showing the contributions of the up (DOS $>$ 0) and down (DOS $<$ 0) spin states in the Cr 3$d$ and I 5$p$ orbitals at (a) 0\,GPa and (b) 25\,GPa. A clear energy gap at 0\,GPa reflects semiconducting character, whereas, the gap disappears at 25\,Gpa signifying metallic character. (c) Pressure-Temperature phase diagram of CrI$_{3}$, reflecting various magnetic and electronic phases. The highest $T_{c}$ corresponds to the calculated Cr-I-Cr bond angle of 89.8$^{\circ}$ at 4\,GPa, which satisfy the GKA rule. The low-temperature and high pressure metallic phase (shaded region) could be the KSL state where the magnetic ordering is highly quenched as pointed out theoretically in\cite{17}.}
\end{figure}

Our spin-polarized DFT also corroborate with the SM transition. In Fig.\,4(a), the conduction band minimum is constituted of spin-polarized up-spin channels of Cr-3$d$ and I-5$p$ states. It can be seen that the band-gap is mainly originated between the I-5$p$ state in the valence band and the Cr-3$d$ state in the conduction band, resulting in a band-gap 0.96 and 2.68\,eV for the spin-up and spin-down electrons, respectively (see Figs.\,S9 and S10). The projected density of states (PDOS) under 25\,GPa, Fig.\,4(b), show a symmetric, unpolarized spin-up and down channels, which also corroborates with the evolution of positive MR above 24\,GPa due to enhanced spin-spin scattering from the non-FM spin orientation of the sample. Our calculations suggest a decrease in the Cr-I-Cr bond angle and Cr-I bond length with pressure. As a consequence, the magnetic ordering is highly suppressed in concurrence with a transition to a metallic phase at high pressure. All these new, hitherto unexplored, high-pressure phases are summarized in the pressure-temperature phase diagram, Fig.\,4(c). At ambient temperature, the resistivity decreases with pressure, resembling a pressure-induced transition to a metallic state. As pointed out in Fig.\,3(d), we mark the metallic region above 22\,GPa. The decrease in resistivity and $T_{c}$ with pressure occur concurrently, meaning that the magnetic ordering (primarily FM) exists in the semiconducting phase, below 21\,GPa. Furthermore, in the metallic regime (above 22\,GPa), the quenching of FM ordering at low temperature reflects a possible KSL state (shaded region in Fig.\,4(c)), as also pointed out in various calculations in \cite{16,30,31}.

Summary and outlook. We have mapped out a pressure-temperature phase diagram of CrI$_{3}$ up to 40\,GPa. We observe, from combined advanced experimental and theoretical investigations, a pressure induced SM transition, that is accompanied by a transition from a robust ferromagnetic state to gradually more dominating anti-ferromagnetic interactions. This makes CrI$_{3}$ a rather unique material, since normally such electronic transitions show exactly the opposite trend (antiferromagnetic interactions turning ferromagnetic). From spin-polarized DFT, we reveal a decrease in the Cr-I-Cr bond angle from 95$^{\circ}$ at ambient pressure to 85$^{\circ}$ at 25\,GPa. The pressure-induced variation in the bond angle affects the FM superexchange interaction and leads to the non-monotonic variation in the $T_{c}$ with pressure. The quenching of ferromagnetic ordering temperature, observed here, represents a possible Kitaev spin-liquid state at low temperature and high pressure, in agreement with recent theories. Our electronic structure calculations support the experimental observation and show the emergence of a finite DOS, primarily contributed by the Cr 3$d$ states, at the Fermi level at and above 25\,GPa. The realization of the KSL state offers tremendous promise for quantum computing and quantum information through the principle of long-range spin-entanglement without the formation of static long-range magnetic ordering in magnetically frustrated systems. The present study will therefore open up new possibilities of extensive research to explore the low-temperature, high-pressure metallic phase, and further calculations of CrI$_{3}$ and similar 2D materials.

A.G. and MAH acknowledges financial support from Carl Tryggers Foundation and the Swedish Research Council (VR) under project No. 2018-05393. G.H. and M.K. acknowledge the support from Czech Science Foundation (project no. 20-08633X). J.V. acknowledges the support of Czech Research Infrastructures MGML (project no. LM2018096). O.E. acknowledge financial support from Knut and Alice Wallenberg Foundation, eSSENCE, SNIC, the Swedish Research Council (VR), the Foundation for Strategic Research (SSF) and the ERC (synergy grant FASTCORR, project 854843). Y.O.K. acknowledges the financial support from the VR under the project No. 2019-03569.


\begin{thebibliography}{100}

\bibitem{1} K. S. Novoselov, A. Mishchenko, A. Carvalho, and A. H. Castro Neto, 2D Materials and van Der Waals Heterostructures, Science \textbf{353}, 6298 (2016).
\bibitem{2}	C. Gong, L. Li, Z. Li, H. Ji, A. Stern, Y. Xia, T. Cao, W. Bao, C. Wang, Y. Wang, Z. Q. Qiu, R. J. Cava, S. G. Louie, J. Xia, and X. Zhang, Discovery of Intrinsic Ferromagnetism in Two-Dimensional van Der Waals Crystals, Nature \textbf{546}, 265 (2017).
\bibitem{3}	M. Bonilla, S. Kolekar, Y. Ma, H. C. Diaz, V. Kalappattil, R. Das, T. Eggers, H. R. Gutierrez, M. H. Phan, and M. Batzill, Strong Room-Temperature Ferromagnetism in VSe2 Monolayers on van Der Waals Substrates, Nat. Nanotechnol. \textbf{13}, 289 (2018).
\bibitem{4}	B. Huang, G. Clark, E. Navarro-Moratalla, D. R. Klein, R. Cheng, K. L. Seyler, Di. Zhong, E. Schmidgall, M. A. McGuire, D. H. Cobden, W. Yao, D. Xiao, P. Jarillo-Herrero, and X. Xu, Layer-Dependent Ferromagnetism in a van Der Waals Crystal down to the Monolayer Limit, Nature \textbf{546}, 270 (2017).
\bibitem{5}	L. Thiel, Z. Wang, M. A. Tschudin, D. Rohner, I. Gutiérrez-Lezama, N. Ubrig, M. Gibertini, E. Giannini, A. F. Morpurgo, and P. Maletinsky, Probing Magnetism in 2D Materials at the Nanoscale with Single-Spin Microscopy, Science \textbf{364}, 973-976 (2019).
\bibitem{6}	N. D. Mermin and H. Wagner, Absence of Ferromagnetism or Antiferromagnetism in One- or Two-Dimensional Isotropic Heisenberg Models, Phys. Rev. Lett. \textbf{17}, 1133 (1966).
\bibitem{7}	O. Besbes, S. Nikolaev, N. Meskini, and I. Solovyev, Microscopic Origin of Ferromagnetism in the Trihalides CrCl$_{3}$ and CrI$_{3}$, Phys. Rev. B \textbf{99}, 104432  (2019).
\bibitem{8}	D. H. Kim, K. Kim, K. T. Ko, J. Seo, J. S. Kim, T. H. Jang, Y. Kim, J. Y. Kim, S. W. Cheong, and J. H. Park, Giant Magnetic Anisotropy Induced by Ligand LS Coupling in Layered Cr Compounds, Phys. Rev. Lett. \textbf{122}, 207201 (2019).
\bibitem{9}	S. W. Jang, M. Y. Jeong, H. Yoon, S. Ryee, and M. J. Han, Microscopic Understanding of Magnetic Interactions in Bilayer CrI$_{3}$, Phys. Rev. Mater. \textbf{3}, 031001(R) (2019).
\bibitem{10}	J. L. Lado and J. Fernandez-Rossier, On the Origin of Magnetic Anisotropy in Two Dimensional CrI$_{3}$, 2D Materials \textbf{4}, 035002 (2017).
\bibitem{11}	A. Frisk, L. B. Duffy, S. Zhang, G. van der Laan, and T. Hesjedal, Magnetic X-Ray Spectroscopy of Two-Dimensional CrI$_{3}$ Layers, Mater. Lett. \textbf{232}, 5 (2018).
\bibitem{12}	J. Kanamori, Superexchange Interaction and Symmetry Properties of Electron Orbitals, J. Phys. Chem. Solids \textbf{10}, 87 (1959).
\bibitem{13}	L. Webster and J. A. Yan, Strain-Tunable Magnetic Anisotropy in Monolayer CrICl$_{3}$, CrBr$_{3}$, and CrI$_{3}$, Phys. Rev. B \textbf{98}, 144411 (2018).
\bibitem{14}	Z. Wu, J. Yu, and S. Yuan, Strain-Tunable Magnetic and Electronic Properties of Monolayer CrI$_{3}$, Phys. Chem. Chem. Phys. \textbf{21}, 7750 (2019).
\bibitem{15}	L. Webster and J. A. Yan, Strain-Tunable Magnetic Anisotropy in Monolayer CrCl$_{3}$, CrBr$_{3}$, and CrI$_{3}$, Phys. Rev. B \textbf{98}, 144411 (2018).
\bibitem{16}	M. Pizzochero and O. V. Yazyev, Inducing Magnetic Phase Transitions in Monolayer CrI$_{3}$ via Lattice Deformations, J. Phys. Chem. C \textbf{124}, 7585 (2020).
\bibitem{17}	S. Mondal, M. Kannan, M. Das, L. Govindaraj, R. Singha, B. Satpati, S. Arumugam, and P. Mandal, Effect of Hydrostatic Pressure on Ferromagnetism in Two-Dimensional CrI$_{3}$, Phys. Rev. B \textbf{99}, 180407 (2019).
\bibitem{18}	See Supplemental Material for the Experimental and Calculation Details and the Supporting Results., n.d.
\bibitem{19}	M. A. McGuire, H. Dixit, V. R. Cooper, and B. C. Sales, Coupling of Crystal Structure and Magnetism in the Layered, Ferromagnetic Insulator Cri3, Chem. Mater. \textbf{27}, 612 (2015).
\bibitem{20}	D. T. Larson and E. Kaxiras, Raman Spectrum of CrI$_{3}$: An Ab Initio Study, Phys. Rev. B \textbf{98}, 085406  (2018).
\bibitem{21}	S. Djurdjic-Mijin, A. $\check{S}$olajic, J. Pe$\check{s}$ic, M. $\check{S}$cepanovic, Y. Liu, A. Baum, C. Petrovic, N. Lazarevic, and Z. V. Popovic, Lattice Dynamics and Phase Transition in CrI$_{3}$ Single Crystals, Phys. Rev. B \textbf{98}, 104307  (2018).
\bibitem{22}	W. Jin, Z. Ye, X. Luo, B. Yang, G. Ye, F. Yin, H. Ho Kim, L. Rojas, S. Tian, Y. Fu, S. Yan, H. Lei, K. Sun, A. W. Tsen, R. He, and L. Zhao, Tunable Layered-Magnetism-Assisted Magneto-Raman Effect in a Two-Dimensional Magnet CrI$_{3}$, Proc. Natl. Acad. Sci. \textbf{117}, 24664 (2020).
\bibitem{23}	P. Nordblad, Disordered Magnetic Systems, Ref. Modul. Mater. Sci. Mater. Eng. 1 (2016) (https://doi.org/10.1016/B978-0-12-803581-8.01101-2).
\bibitem{24}	X. Ke, M. L. Dahlberg, E. Morosan, J. A. Fleitman, R. J. Cava, and P. Schiffer, Magnetothermodynamics of the Ising Antiferromagnet Dy$_{2}$Ge$_{2}$O$_{7}$, Phys. Rev. B \textbf{78}, 104411 (2008).
\bibitem{25}	D. R. Klein, D. MacNeill, J. L. Lado, D. Soriano, E. Navarro-Moratalla, K. Watanabe, T. Taniguchi, S. Manni, P. Canfield, J. Fernández-Rossier, and P. Jarillo-Herrero, Probing Magnetism in 2D van Der Waals Crystalline Insulators via Electron Tunneling, Science \textbf{360}, 1218 (2018).
\bibitem{26}	T. Song, X. Cai, M. W. Y. Tu, X. Zhang, B. Huang, N. P. Wilson, K. L. Seyler, L. Zhu, T. Taniguchi, K. Watanabe, M. A. McGuire, D. H. Cobden, D. Xiao, W. Yao, and X. Xu, Giant Tunneling Magnetoresistance in Spin-Filter van Der Waals Heterostructures, Science \textbf{360}, 1214-1218 (2018).
\bibitem{27}	K. Sato and E. Saitoh, Spintronics for Next Generation (John Wiley and Sons, Ltd, 2015).
\bibitem{28}	A. Kitaev, Anyons in an Exactly Solved Model and Beyond, Ann. Phys. (N. Y). \textbf{321}, 2 (2006).
\bibitem{29}	R. Yadav, S. Rachel, L. Hozoi, J. Van Den Brink, and G. Jackeli, Strain- and Pressure-Tuned Magnetic Interactions in Honeycomb Kitaev Materials, Phys. Rev. B \textbf{98}, 121107(R) (2018).
\bibitem{30}	I. Lee, F. G. Utermohlen, D. Weber, K. Hwang, C. Zhang, J. Van Tol, J. E. Goldberger, N. Trivedi, and P. C. Hammel, Fundamental Spin Interactions Underlying the Magnetic Anisotropy in the Kitaev Ferromagnet CrI$_{3}$, Phys. Rev. Lett. \textbf{124}, 17201 (2020).
\bibitem{31}	C. Xu, J. Feng, M. Kawamura, Y. Yamaji, Y. Nahas, S. Prokhorenko, Y. Qi, H. Xiang, and L. Bellaiche, Possible Kitaev Quantum Spin Liquid State in 2D Materials with S=3/2, Phys. Rev. Lett. \textbf{124}, 87205 (2020).



\end{thebibliography}
\end{document}